\documentclass[12pt]{iopart}
\usepackage{iopams}
\usepackage{graphicx}
\usepackage{dcolumn}
\usepackage{bm}
\newcommand{\fracc}[2]{\frac{\textstyle{#1}}{\textstyle{#2}}}
\newcommand{\pra}{{\it Phys.\ Rev.\ A}}

\newcommand{\prc}{{\it Phys.\ Rev.\ C}}
\newcommand{\prd}{{\it Phys.\ Rev.\ D}}
\newcommand{\pre}{{\it Phys.\ Rev.\ E}}
\newcommand{\apj}{{\it Astrosph.\ J.}}
\begin{document}

\title{The flexibility of optical metrics}

%\author{Eduardo Bittencourt$^{1,2}$}
%\email{eduardo.bittencourt@icranet.org}
%\author{Jonas P. Pereira$^{1,3}$}
%\email{jpereira@towson.edu}
%\author{Igor I. Smolyaninov$^{4}$}
%\email{smoly@umd.edu}
%\author{Vera N. Smolyaninova$^{3}$}
%\email{vsmolyaninova@towson.edu}
%\affiliation{$^{1}$CAPES Foundation, Ministry of Education of Brazil, Bras\'ilia, Brazil}
%\affiliation{$^{2}$Sapienza Universit\`a di Roma - Dipartimento di Fisica, P.le Aldo Moro 5 - 00185 Rome - Italy}
%\affiliation{$^{3}$Department of Physics, Astronomy and Geosciences, Towson University, 8000 York Road, Towson, Maryland 21252-0001, USA}
%\affiliation{$^{4}$Department of Electrical and Computer Engineering, University of Maryland, College Park, MD 20742, USA }

\author{Eduardo Bittencourt$^{1,2}$, Jonas P. Pereira$^{1,3}$, Igor I. Smolyaninov$^{4}$, Vera N. Smolyaninova$^{3}$}
\address{$^{1}$CAPES Foundation, Ministry of Education of Brazil, Bras\'ilia, Brazil}
\address{$^{2}$Sapienza Universit\`a di Roma - Dipartimento di Fisica, P.le Aldo Moro 5 - 00185 Rome - Italy}
\address{$^{3}$Department of Physics, Astronomy and Geosciences, Towson University, 8000 York Road, Towson, Maryland 21252-0001, USA}
\address{$^{4}$Department of Electrical and Computer Engineering, University of Maryland, College Park, MD 20742, USA }
\eads{\mailto{eduardo.bittencourt@icranet.org}, \mailto{jpereira@towson.edu}, \mailto{smoly@umd.edu}, \mailto{vsmolyaninova@towson.edu}}

\date{\today}

\begin{abstract}
We firstly revisit the importance, naturalness and limitations of the so-called optical metrics for describing the propagation of light rays in the limit of geometric optics. We then exemplify their flexibility and nontriviality in some nonlinear material media and in the context of nonlinear theories of the electromagnetism, both in presence of curved backgrounds, where optical metrics could be flat and inaccessible regions for the propagation of photons could be conceived, respectively. Finally, we underline and discuss the relevance and potential applications of our analyses in a broad sense, ranging from material media to compact astrophysical systems.
\end{abstract}

\maketitle

%%%%%%%%%%%%%%%%%%%%%%%%%%%%%%%%%%%%%%%%%%%%%%%%%%%%%%%%%%%%%%%%%%%%%%%%%%%%%
%%%%%%%%%%%%%%%%%%%%%%%%%%%%%%%%%%%%%%%%%%%%%%%%%%%%%%%%%%%%%%%%%%%%%%%%%%%%%
\section{Introduction}
%%%%%%%%%%%%%%%%%%%%%%%%%%%%%%%%%%%%%%%%%%%%%%%%%%%%%%%%%%%%%%%%%%%%%%%%%%%%%
%%%%%%%%%%%%%%%%%%%%%%%%%%%%%%%%%%%%%%%%%%%%%%%%%%%%%%%%%%%%%%%%%%%%%%%%%%%%%
It is very well-known that light propagation in the limit of geometric optics can be described by Fermat's principle
\cite{1960ecm..book.....L, 1999poet.book.....B}. It states that the trajectories of rays can be obtained by the extremization of
their optical paths \cite{1999poet.book.....B}. Geodesics in a spacetime are obtained likewise, by extremizing the distance between
two given spacetime events. This means that these two issues are intimately related (\cite{1999poet.book.....B},
page 127, and references therein) and aspects raising in one area should be obtained in the other. This must be specially the case for
the spacetime metric. Indeed, it has been already shown how this geometric structure steering light propagation always arise in the
context of nonlinear theories of the electromagnetism
\cite{1970JMP....11..941B, 2000PhRvD..61d5001N, 2000PhRvA..62a2111L, 2001PhRvD..63j3516N, 2002PhRvD..65f3501D, 2002PhRvD..66b4042O, 2007CQGra..24.3021N},
some linear \cite{1923AnP...377..421G, 1999PhLB..458..466O, 2000PhRvD..62d4050O, 2003CQGra..20..859N} and nonlinear media
\cite{2001PhLB..512..417D, 2002PhRvE..65b6612D, 2002PhRvD..65f4027D, 2003PhRvD..68f1502D, 2010NJPh...12i5021C}, condensed matter (\cite{2011LRR....14....3B} and
references therein), etc.
Approaches where effective metrics arise are generically called analogue models
\cite{1981PhRvL..46.1351U,2011LRR....14....3B, 2011CQGra..28n5022N, 2008arXiv0805.4778L} (mainly in the optical and acoustic scopes) and they have been studied in a variety of scenarios, such as Hawking radiation coming from analogue apparent horizons \cite{2003IJMPD..12..649V, 2010PhRvL.105t3901B,2011PhRvL.107n9401S, 2008Sci...319.1367P,2008NJPh...10e3015R, PhysRevLett.106.021302, 2012CQGra..29v4009F} and analogue quantum gravity (see e.g. \cite{2010PhRvA..82e3602K,2011LRR....14....3B,2014PhRvD..90j4015B,2007arXiv0712.0427V} and references therein), only to cite some.
It should be stressed that in these models the concept of optical metrics (in this work we shall not focus on acoustic metrics), effective metrics for the propagation of light rays, is an emergent aspect.
Once established, they can be regarded as fundamental quantities, for they directly describe photon trajectories.
This is to be contrasted with the metric of the background spacetime, which is a fundamental quantity present ab initio in the Lagrangian density of interest,
related to what optical metrics shall also depend upon. With the use of the equivalence principle, optical geometries can be defined in any background spacetime.

The control of light trajectories can be approached, for instance, with transformation optics (see Ref. \cite{2008arXiv0805.4778L} for a comprehensive review and applications).
It is important since it takes advantage of the symmetries of Maxwell equations in vacuum and in continuous media to make a link between coordinate transformations (or geometries) and dielectric coefficients \cite{1975ctf..book.....L}. Nonetheless, this analogy is only valid when the permittivity and permeability tensors are equal, which is very restrictive in terms of generic dielectric tensors and, in general, can only be attained with metamaterials
\cite{2000PhRvL..84.4184S, 2001Sci...292...77S, 2010PhRvL.105f7402S}.

In the context of nonlinear theories of the electromagnetism, it is always possible to obtain effective metrics out of propagation of light rays. This means that given a Lagrangian density dependent upon the invariants of the electromagnetism, one can always ``geometrize'' the propagation of photons in terms of it; see \cite{2002PhRvD..66b4042O} and references therein.
Nevertheless, this is not generally the case for light propagation in material media and for linear ones it is already known that a closure relation should be satisfied by the dielectric coefficients (permittivity and permeability) such that emergent optical metrics be dependent only upon spacetime coordinates \cite{2000PhRvD..62d4050O}. A generalization to the closure relation does not yet exist for nonlinear media, in spite of the fact that optical geometries arise for some nonlinear isotropic liquids \cite{2001PhRvD..63j3516N,2008PhRvD..78d5015D}. Without the closure relation an optical metric would generically depend upon properties of photons, such as their wave-vectors. This would be of interest, for instance, to investigate extensions of Riemannian geometries, like Finsler models, shedding some light into quantum gravity phenomenology, due to modified energy-momentum dispersion relations, or the dark matter problem (\cite{2013LRR....16....5A,1993PhRvD..48.3641B} and references therein).

One aspect related to effective geometries that is not usually recalled is that, due to the fact they depend upon external electromagnetic fields and dielectric coefficients, for instance, in principle it is possible to conceive scenarios where they are very simple (such as Minkowskian or conformally Minkowskian), even in curved background spacetimes, when the aforesaid fields and dielectric tensors could be controlled or presented convenient properties.
This leads precisely to the novelty of this work: to show the significance, physical reasonableness and
nontriviality of the cases where optical metrics are rendered or naturally are simple. Such an analysis could have far-reaching implications for material sciences, charged black hole spacetimes, compact systems, etc. For instance, as we shall underline and exemplify, flat (Minkowski) optical metrics for material media would lead to the practical absence of reflection and refraction of light rays coming from/going to an almost vacuous region (since our analyses will be done in the limit of geometric optics and for usual media, one can assume that their permeabilities are very close to that of vacuum [nonmagnetic media] \cite{2010opme.book.....C,1999poet.book.....B,1960ecm..book.....L}, which in practical terms means that rays do not reflect if the medium they come from has the same index of refraction as the one they go to); for compact systems, it would lead photons to propagate freely there. We stress that we are not interested here in elaborating on specific analogue models of gravity, but rather to discuss the relevance and pertinence of cases more related to its absence when curved background spacetimes are taken into account.

This article is organized as follows. In section \ref{optical metric} we revisit how the geometric interpretation may arise and its limitations for light propagation in the limit of geometric optics. Section \ref{exploring} is devoted to the study (for the first time) of some cases where optical metrics could be Minkowski or almost Minkowski in the contexts of charged black holes described by nonlinear Lagrangians of the electromagnetism, nonlinear liquid media and in some regions of neutron stars. This could happen if some of their parameters fulfilled some constraints, which we derive and discuss in detail, specially regarding their physical reasonableness. In section \ref{conc} we gather the key points raised and discuss possible far-reaching consequences of our analyses for material media and astrophysics.

Units here are such that $c=G=1$, unless otherwise stated.

%%%%%%%%%%%%%%%%%%%%%%%%%%%%%%%%%%%%%%%%%%%%%%%%%%%%
%%%%%%%%%%%%%%%%%%%%%%%%%%%%%%%%%%%%%%%%%%%%%%%%%%%%
\section{Effective geometries for light propagation revisited}
\label{optical metric}
%%%%%%%%%%%%%%%%%%%%%%%%%%%%%%%%%%%%%%%%%%%%%%%%%%
%%%%%%%%%%%%%%%%%%%%%%%%%%%%%%%%%%%%%%%%%%%%%%%%%%%
In the limit of geometric optics, the important concept is that of rays, in very close resemblance with
material particles \cite{1960ecm..book.....L}. Nevertheless, wave properties are also important, since in the geometric optics limit wave packets
could always be made, leading to ray description. Let us start with a monochromatic plane
wave described by its wave four-vector $k_{\mu}=(\omega, -\vec{q})$, where $\omega$ is its frequency and $\vec{q}$
its wave vector that defines the direction of propagation. In the limit of geometric optics, such a  four-vector
must be orthogonal to a hypersurface [the eikonal $\Psi(x^{\nu})$], thus $k_{\mu}\doteq\partial_{\mu}\Psi$.

Let us assume that the background spacetime is arbitrary. Making use of a locally inertial frame, the
equations governing the propagation of rays are \cite{1960ecm..book.....L, 1975ctf..book.....L}
\begin{equation}
\frac{d\vec{q}}{d\bar{t}}=-\frac{\partial \omega}{\partial \vec{r}}, \qquad \frac{d\vec{r}}{d\bar{t}}=\frac{\partial \omega}{\partial \vec{q}}\label{geoopteq},
\end{equation}
where $\bar t$ is the time recorded there and $\omega= \omega(\vec{r},\vec{q},\bar{t})$. The latter relation is obtained by dint of the Fresnel equation (dispersion relation) \cite{1960ecm..book.....L,1999poet.book.....B} to the medium under interest. For media whose dielectric coefficients are functions of the space-time coordinates or for those ones that are frequency dependent, but operating far from resonance or plasma frequencies, the Fresnel equation can always be written as \cite{1970JMP....11..941B, 2000PhRvD..62d4050O, 2002PhRvD..66b4042O}
\begin{equation}
\widehat{G}^{\mu \nu \alpha \beta}(x^{\rho}) k_{\mu} k_{\nu} k_{\alpha} k_{\beta}=0\label{Fresnelgen},
\end{equation}
where $\widehat{G}^{\mu \nu \alpha \beta}$ is a rank-4 completely symmetric tensor depending only on the electromagnetic field and its continuous derivatives.
In this work we shall be concerned with situations in which $\hat{G}^{\mu\nu\alpha\beta}$ is decomposable in terms of a tensorial product of $\hat{g}^{\mu\nu}$s (naturally occurring in nonlinear Lagrangians of the electromagnetism \cite{2002PhRvD..66b4042O} and for some nonlinear isotropic media \cite{2001PhRvD..63j3516N,2008PhRvD..78d5015D}), such that Eq.\ (\ref{Fresnelgen}) can be factorized in quadratic terms of the form
\begin{equation}
\Upsilon\equiv \hat g^{\alpha\beta} (x^{\mu}) k_{\alpha} k_{\beta}=0\label{Fresnelrestrict}.
\end{equation}
We shall see that only in such situations one could make analogies of the propagation of light rays and geodesics in a curved manifold.

Given Eq.\ (\ref{Fresnelrestrict}), it is simple to show that Eqs.\ (\ref{geoopteq}) can be cast as
\cite{1970JMP....11..941B}
\begin{equation}
\frac{dx^{\beta}}{d\sigma}=\frac{\partial \Upsilon}{\partial k_{\beta}}, \qquad \frac{dk_{\beta}}{d\sigma}=-\frac{\partial \Upsilon}{\partial x^{\beta}}\label{geooptcovariant},
\end{equation}
where we have defined $d\sigma \doteq d\bar{t} \partial \omega/\partial \Upsilon$ as a parameter along the ray trajectories.
Clearly Eq.\ (\ref{geooptcovariant}) is the covariant version of Eqs.\ (\ref{geoopteq}) and therefore is valid in any coordinate system.

In order to obtain the ray paths, one has to solve Eqs.\ (\ref{geooptcovariant}). For Eq.\ (\ref{Fresnelrestrict}) this solution can be obtained generically, as we shall show now. From Eqs.\ (\ref{geooptcovariant}) and
Eq.\ (\ref{Fresnelrestrict}), we have that
\begin{equation}
\frac{dx^{\alpha}}{d\sigma}\doteq \dot{x}^{\alpha}= 2\hat g^{\alpha\beta}k_{\beta}\rightarrow k_{\mu}=\frac{1}{2}\hat g_{\mu\alpha}\dot{x}^{\alpha}, \label{Kintermsofraypath}
\end{equation}
where $\hat g_{\mu\nu}$ is defined such that $\hat g_{\mu\nu} \hat g^{\mu\beta} = \delta^{\beta}{}_{\nu}$.
By differentiating $k_{\mu}$ in the above equation with respect to $\sigma$ and using the second equation in Eq.\ (\ref{geooptcovariant}), after some simplifications one has that
\begin{equation}
\ddot{x}^{\nu} + \frac{1}{2}\hat g^{\mu\nu}(\hat g_{\mu\alpha,\beta} + \hat g_{\mu\beta,\alpha} - \hat g_{\alpha\beta,\mu}) \dot{x}^{\alpha} \dot{x}^{\beta}=0 \label{ddotx}.
\end{equation}
Therefore, if one assumes that rays propagate in an effective spacetime whose geometrical structure is encompassed by $\hat g_{\mu\nu}$,
then the second term of the left-hand side of Eq.\ (\ref{ddotx}) is proportional to the Christoffel symbols.
Based on this effective geometric structure, Eq. (\ref{ddotx}) can be cast as
\begin{equation}
\ddot{x}^{\mu}+ \widehat\Gamma^{\mu}_{\alpha\beta} \dot{x}^{\alpha} \dot{x}^{\beta}=0 \label{ddotgeodesic},
\end{equation}
which makes it clear that rays follow along null geodesics in an effective spacetime metric $\hat g_{\mu\nu}$ [according to the definition of $\Upsilon$ and Eqs.\ (\ref{geooptcovariant})]:
\begin{equation}
\dot{x}^{\alpha}k_{\alpha}=0\,\longleftrightarrow\, \hat g_{\mu\nu}\dot{x}^{\mu}\dot{x}^{\nu}=0\label{nullfourvelocity}.
\end{equation}
Therefore, in order to know the ray trajectories, the effective metric structure is vital. Note that light-cones are determined by $\hat g_{\mu\nu}$  [see the above equation]. Out of the limit of geometric optics, the background metric is the responsible for the causal structure.

Let us now say a few words about the case where $\hat{g}^{\mu\nu}= \hat{g}^{\mu\nu}(x^{\beta},k_{\alpha})$. In such a scenario Eq.\ (\ref{Kintermsofraypath}) will be modified and by consequence the right-hand side of Eq.\ (\ref{ddotgeodesic}) will gain terms not proportional to $\dot{x}^{\mu}$. Therefore, only the case $\hat{g}^{\mu\nu}=\hat{g}^{\mu\nu}(x^{\beta})$ is connected with usual geodesics for light rays.

Concerning the geometrization of light propagation in material media, we point out that the approach we made use of is slightly different from the one coming from transformation optics. The former deals with given dielectric coefficients (so there is no induction of a material medium due to a coordinate transformation), which by means of the Fresnel equation are mapped onto induced geometries for light rays (either ordinary or extraordinary). For light propagation in a nonlinear theory of the electromagnetism, one can indeed map it onto a continuous dielectric medium, but the associated permittivity and permeability tensors are generally nonlinear and different \cite{1934RSPSA.144..425B, 2014PhRvA..89d3822D}. Specific properties of light rays in our description arise from convenient optical metrics (or effective geometries), similarly as in transformation optics.

\section{Exploring effective metrics}
\label{exploring}
Contrary to the usual cases in which nonlinear theories of electromagnetism in flat space in the limit of geometric optics lead to the appearance of an effective curved metric for the propagation of perturbations, in this paper we wonder under which circumstances some nonlinear theories defined in a curved spacetime could cancel out the effects of gravity, such that perturbations would follow along light-like curves of Minkowski spacetime. Similarly, one can also wonder the properties nonlinear media and fields in given curved backgrounds should have so that their optical metrics be flat. Thus, in this section we scrutinize the properties some well-known electromagnetic systems in the literature (but very different in their nature) should have in order that their optical metric structure be simple (such as Minkowskian).

\subsection{Controlling the effective metric of nonlinear black hole spacetimes}
Here we shall investigate the conditions a nonlinear Lagrangian, an electric field and a background metric (all of them connected by the field equations and not fixed a priori) should satisfy such that the associated optical metric be ``controllable''. It is already known that for any Lagrangian $L(F)$ (valid for configurations where the magnetic field is absent), $F\doteq g^{\alpha\mu}g^{\beta\nu}F_{\alpha\beta}F_{\mu\nu}$ is the invariant constructed from the electromagnetic tensor $F_{\mu\nu}$, the following effective metric emerges for the propagation of disturbances in this
context \cite{2000PhRvD..61d5001N}:
\begin{equation}
\hat g^{\mu\nu}\doteq g^{\mu\nu} + 4\frac{L_{FF}}{L_F}g_{\alpha\beta}F^{\mu\alpha}F^{\nu\beta}\label{HmunuNLED},
\end{equation}
where $L_{F}\doteq \partial L/\partial F$ and $L_{FF}$ is instead the second derivative of $L$ with respect to $F$.

To illustrate what may happen with nonlinear electromagnetic theories in curved backgrounds, let us analyze the case of a spherically symmetric nonlinear charged black hole related to a nonlinear Lagrangian $L(F)$, in the realm of general relativity. It is simple to show that its background metric is given by (see for instance \cite{2014PhLB..734..396P})
\begin{equation}
\frac{d(g_{00}r)}{dr}=1-2QE_r+2Lr^2\label{g00NLED},
\end{equation}
where $Q$ is a constant of integration (charge of the system) and the electric field $E_r$ itself is determined by
\begin{equation}
\frac{d L}{d E_r}= \frac{Q}{r^2}\label{ErNLED},
\end{equation}
which is nothing more than the sole equation of the electromagnetism for the case under interest coming from an arbitrary $L(F)$, given that there $F=-2E_r^2$. If the nonlinear Lagrangian to the system is known, then the effective geometry is given and only its consequences could be probed.

For the problem of a nonlinear black hole spacetime, in order to have a controllable effective metric, strictly speaking, one must assume that the nonlinearities of the Lagrangian are to be fixed, as well as the electric field of the system. This case may be of relevance in assessing certain kinds of nonlinearities of the electromagnetism whose associated light propagation in the corresponding nonlinear black hole spacetimes are simple, irrespective of how involved the background metrics may be. Let us start our analyses by assuming that
\begin{equation}
1-\frac{4L_{FF}}{L_F}E_r^2\equiv f(r)\label{geffcontraint},
\end{equation}
where $f(r)$ is a given function of the radial coordinate that satisfies the requirement $f(\infty)=1$. From Eq. (\ref{HmunuNLED}) one clearly sees this is the sole term related to the electrodynamics of the system that contributes to its effective geometry. Simplifications of Eq.\ (\ref{geffcontraint}) taking into account Eq. (\ref{ErNLED}) leads to the generic solution
\begin{equation}
E_r(r)= Q\exp\left[-2\int \frac{dr}{f(r)r}\right]\label{Ersolution},
\end{equation}
where we considered that its asymptotic solution is Minkowskian, as implied by $f(\infty)=1$.

Let us write the Lagrangian density as
\begin{equation}
L=\frac{E_r^2}{2}+L_{nl}\label{L},
\end{equation}
where $L_{nl}$ is a nonlinear term and $E_r^2/2$ corresponds to the Maxwellian contribution for the Lagrangian. From Eqs. (\ref{ErNLED}) and (\ref{L}), one shows that $L_{nl}$ must satisfy the condition
\begin{equation}
\frac{dL_{nl}}{dr}= \frac{dE_r}{dr}\left(\frac{Q}{r^2}-E_r \right)\label{Lnl},
\end{equation}
with the constraint that $L_{nl}(\infty)=0$. Therefore, the hierarchy of the problem under interest is the following: choose a $f(r)$ such that $f(\infty)=1$; then find the electric field $E_r$ by solving Eq.\ (\ref{Ersolution}); afterwards solve Eq.\ (\ref{Lnl}) and, finally, solve for $g_{00}$ according to Eq.\ (\ref{g00NLED}). For each choice of $f(r)$, the effective metric becomes
\begin{equation}
ds^2_{\rm{eff}}= \frac{1}{f(r)}\left[g_{00}\,dt^2- \frac{dr^2}{g_{00}} \right]-r^2d\theta^2-r^2\sin^2\theta d\varphi^2
\label{ds2eff},
\end{equation}
while the background metric is
\begin{equation}
ds^2= g_{00}\,dt^2- \frac{dr^2}{g_{00}} -r^2d\theta^2-r^2\sin^2\theta d\varphi^2\label{ds2}.
\end{equation}

For the case of black holes, it would be of interest to analyze the situation where $f(r)=g_{00}$, since it would lead to a simple optical metric structure.
In this case, from Eqs. (\ref{g00NLED}), (\ref{Ersolution}) and (\ref{Lnl}), one shows that for $E_r$ different from a constant
\begin{eqnarray}
3(E_{r,r})^3&+&E_rE_{r,r}[rE_{r,r^3}-2E_{r,r^2}-2Q(E_{r,r})^2]\nonumber \\&+& rE_{r,r^2}[(E_{r,r})^2-2E_r E_{r,r^2}] =0\label{Erfg00}
\end{eqnarray}
and (background metric)
\begin{equation}
g_{00}= -\frac{2E_r}{rE_{r,r}}\label{g00fg00},
\end{equation}
where we have defined $E_{r,r^n}\doteq d^nE_r/dr^n$.
In what follows we shall not investigate Eqs. (\ref{Erfg00}) and (\ref{g00fg00}) but rather make generic analyses of photon propagation. We start by focusing on the role played by $f(r)$ on radial geodesics of the effective metric, in particular at the points where $g_{00}$ is null, which invalidate the conformal analogy between (\ref{ds2eff}) and (\ref{ds2}). Apparent singularities of the background metric (\ref{ds2}) can be removed by introducing the Painlev\'e-Gullstrand (PG) coordinates, defined as $T=t+\int(\sqrt{1-g_{00}}/g_{00})dr$. The integration of the ingoing null radial geodesics related to Eq.\ (\ref{ds2eff}) in the aforesaid coordinates leads us to
\begin{eqnarray}
&\dot r=-f,\label{rdot}\\[1ex]
&\dot T=\fracc{f}{g_{00}}\left(1-\sqrt{1-g_{00}}\right),\label{tdot}
\end{eqnarray}
where $\dot X\equiv dX/d\sigma$ and $\sigma$ is an affine parameter. For the background spacetime [$f(r)=1$], the above equations show that ingoing light rays cross its outer horizon (denoted from now on by $r_+$). Nothing similar happens with the effective metric when $f=g_{00}$. From Eqs.\ (\ref{rdot}) and (\ref{tdot}), one can easily see that in this case $\ddot r=\dot r=\ddot T=\dot T=0$ at $r=r_+$, indicating that light rays cannot cross from outside to inside of the outer horizon, since external observer fields do not perceive any singularity there. In other words, the outer horizon avoids that any eventual singularity in the background metric can be reached by photons. A direct way of seeing that indeed photons cannot cross $r=r_+$ is by introducing the new ``radial'' coordinate $R=\int(dr/g_{00})$, which leads Eq.~(\ref{ds2eff}) to
\begin{equation}
ds^2_{\rm{eff}}= dt^2- dR^2-r(R)^2(d\theta^2+\sin^2\theta d\varphi^2).
\label{ds2eff2}
\end{equation}
By assuming that $g_{00}$ is a rational function that vanishes (at least) at $r=r_+$, one has that the domain of $R$ is the whole real line, while its image $r=r(R)$ is bound (from below) by $r=r_+$. This indicates that the region $r<r_+$ is excluded from the portion of the space-time accessible to the ingoing photons; in this coordinate system, radial null geodesics are {\it maximally extended}. Note also from Eq.\ (\ref{ds2eff2}) that in the case under investigation it is impossible to render the effective metric Minkowskian.

\subsection{Minkowskian extraordinary rays in nonlinear liquids in curved backgrounds}
We shall now analyze the case of nonlinear media in the context of curved backgrounds such that its optical metric be flat. In particular, we shall use liquid media since under natural circumstances they are isotropic. This simplifies tremendously the propagation of disturbances there \cite{2008PhRvD..78d5015D}. When electromagnetic fields are present in these media, they might eventually become optically anisotropic (leading to the phenomenon of birefringence), with the induced optical axis reflecting their symmetries \cite{2008PhRvD..78d5015D}.

Just in order to be simple and to evidence the physical ideas we want to convey, let us analyze Kerr media (see \cite{2008PhRvD..78d5015D} and references therein). For this case, we have that its permeability is a constant, defined by $\mu_0$, and its permittivity is given by $\varepsilon^{\alpha}{}_{\beta}= \epsilon(E) (\delta^{\alpha}{}_{\beta}-V^{\alpha}V_{\beta})$, where $E=\sqrt{-E^{\mu}E_{\mu}}$ is the norm of the external electric field $E^{\mu}$ and $V^{\alpha}$ the four-velocity of an observer, orthogonal to $E^{\mu}$, who measures the electric field. Let us investigate here only extraordinary photons, related to the anisotropic solution of the Fresnel equation in the presence of electric fields. Ordinary photons will be generically described at the end of this section. The associated effective metric for this case can be obtained from its Fresnel equation and the result, when generalized to an arbitrary coordinate system, is \cite{2001PhRvD..63h3511N}
\begin{equation}
\hat g^{\mu\nu}= g^{\mu\nu} - \left[1-\mu_0(\epsilon + \epsilon'E^2)\right]V^{\mu}V^{\nu} - \frac{\epsilon'}{\epsilon}E^{\mu}E^{\nu}\label{Kerrmetric},
\end{equation}
with $\epsilon'\doteq (1/E)\partial \epsilon/ \partial E$.
Observe that for the case $\epsilon$ is a constant, Eq.\ (\ref{Kerrmetric}) gives us the well-known Gordon metric (see for instance Refs. \cite{1923AnP...377..421G, 2012PhRvD..86l4024N}).
It is simple to show that the inverse of Eq. (\ref{Kerrmetric}) is
\begin{equation}
\hat g_{\mu\nu} = g_{\mu\nu} +\frac{1-\mu_0(\epsilon + \epsilon'E^2)}{\mu_0(\epsilon + \epsilon'E^2)}V_{\mu}V_{\nu} +
\frac{\epsilon'}{(\epsilon + \epsilon'E^2)}E_{\mu}E_{\nu}\label{HmunuKerr}.
\end{equation}
Assume a (given) background metric with spherical symmetry such that $g_{\mu\nu}=\rm{diag}(g_{00},g_{11},-r^2,-r^2\sin^2\theta)$. Consider the same symmetry for the (unknown) electric field, that with respect to an observer at rest [$V_{\mu}=\sqrt{g_{00}}\,\delta_{\mu}^0$] is then $E^{\mu}= E_r(r)\delta^{\mu}_1$. For this case, Eq.\ (\ref{HmunuKerr}) admits the solution $\hat g_{\mu\nu}=\eta_{\mu\nu}$ as far as
\begin{equation}
g_{00}=\mu_0(\epsilon +\epsilon' E^2)\label{h00cond1}
\end{equation}
and
\begin{equation}
\frac{\epsilon'\,E^2}{\epsilon}= -(g_{11}+1)\label{h11cond2}.
\end{equation}

When Eq. (\ref{h11cond2}) is inserted into Eq. (\ref{h00cond1}), one has that
\begin{equation}
\mu_0\epsilon=\frac{g_{00}}{(-g_{11})}\label{mu_epsilon}.
\end{equation}
From the radial symmetry involved and Eqs. (\ref{h11cond2}) and (\ref{mu_epsilon}), it follows that the generic solution to the norm of the electric field can be cast as
\begin{equation}
E(r)=E_0\exp\left[-\int_{r_0}^{r}\frac{1}{1+g_{11}}\frac{\partial}{\partial \bar{r}}\log\left(-\frac{g_{00}}{g_{11}} \right) d\bar{r} \right]\label{Er},
\end{equation}
where $E_0=E(r_0)$.
In order to exemplify the above formalism, let us investigate the case $g_{11}=-1/g_{00}$, valid, for instance, for exterior solutions to general relativity in the spherically symmetric case. Here, Eq. (\ref{Er}) simplifies to
\begin{equation}
E(r)= E_0\left[\frac{g_{00}(r_0)-1}{g_{00}(r)-1}\right]^2\label{Eg00}.
\end{equation}
For a concrete example, assume $g_{00}=1-2M/r$, where $M$ is a constant and $r>2M$. Then, Eq.~(\ref{Eg00}) reads
\begin{equation}
E(r)=E_0\left(\frac{r}{r_0} \right)^2\label{ErSchw}.
\end{equation}
From Maxwell equations in material media, given that $E=\sqrt{-g_{11}}E_r$, the associated charge density is
\begin{equation}
\rho(r)=\frac{1}{4\pi r^2}\frac{\partial}{\partial r}\left[r^2\epsilon E_r\right]= \frac{E_0}{4\pi\mu_0}\frac{\left(r-{2M}\right)^{\frac{3}{2}}\left(4r-3M\right)}{r_0^2}\label{charge_density}.
\end{equation}
One clearly sees from Eqs. (\ref{ErSchw}) and (\ref{charge_density}) that the electric field is never null and that the system has a nonzero net charge. For situations where $2M/r\ll 1$, for instance, one has from the above equations that $\mu_0\epsilon \approx 1-4M/r$, which means $\mu_0\epsilon(E) \approx 1- 4(M/r_0)\sqrt{E_0/E}$. This is the dielectric response a Kerr liquid medium must have in the presence of the electric field with norm given by Eq. (\ref{ErSchw}) for supporting straight light ray trajectories with speed $c$ there.

We finally remark that all the analyses in this section were related to the optical metric felt by extraordinary rays. The ordinary one is simply given
by a Gordon-like metric, obtained from Eq. (\ref{HmunuKerr}) by assuming $\epsilon'\rightarrow 0$. As it can be checked in the particular example of this section, it is not possible to render both ordinary and extraordinary optical metrics Minkowskian at the same time, since they are very different in their functional forms. Generally, one could in principle only render the ordinary metric Minkowskian in moving media \cite{1999PhRvA..60.4301L}, since
further degrees of freedom should emerge there.

\subsection{Flat optical metrics in compact astrophysical systems?}
One could also imagine astrophysical consequences related to the flexibility of optical metrics. Consider for instance the photon propagation in the interior of compact objects. Such photons are present in material media, not in vacuum, and therefore in the limit of geometric optics it would also be necessary to know the dielectric properties of the astrophysical system in order for photons to be properly described. Photon trajectories {\it must} be related to effective geometries and they carry the information of the underlying spacetime metric and medium.

Naively speaking, one would expect that the presence of a material medium would considerably slow the photons and all the information we can get from the astrophysical system would come from the outermost regions. Notwithstanding, we have seen that whenever an optical metric becomes flat photons would not be trapped anymore and could freely escape the system. This could constitute a mechanism to release energy from compact stars and coalescent systems that could take place intermittently, due generally to their nontrivial dynamical evolution.

Let us analyze, only as a particular case of the above discussion, the conditions required for the extraordinary optical metric to be Minkowskian in the vicinities of the origin of a static, charged and spherically symmetric compact system. Such an analysis is deemed to be important since it would give us insights into the physically interesting cases to pay attention in more precise investigations.
The main point here is not related to the strength of the nonlinearities in the dielectric coefficients (obviously small for regular electric fields), but solely to their existence, since it is only in this case they would naturally be connected with electric fields, which must coincide with the ones generated by charge distributions in stars and, thus, be regular. Analyses of this sort are motivated by the possibility of allowing photons produced in the innermost regions of compact stars to propagate freely outwards, giving us hope to assess how such regions are. In any other case this seems highly improbable.

Near the center of the aforesaid system, the total energy density $\rho$ (taking into account the contributions due to electromagnetic and matter fields) could always be Taylor-expanded, yielding
\begin{equation}
\rho\approx \rho_0-\sum_{n=1}^{N}\beta_{n} r^{n}\label{rho},
\end{equation}
where $\beta_{n}$ and $N\in\mathbb{N}$ are constants, and $\rho_0$ is the total energy density at the origin. Note that some $\beta_n$'s are null, such as the ones related to odd $n$, due to symmetry criteria. Nevertheless, by making use of last subsection's results, we shall show that their specification is irrelevant for the regularity of an electric field in a finite region around the origin of a star (by choosing a convenient $N$ for so). Let us define, only for convenience, $g_{00}\doteq e^{\nu}$ and $g_{11}\doteq -e^{\lambda}$. In the case we are interested in, Einstein equations lead to \cite{1975ctf..book.....L}
\begin{equation}
e^{-\lambda}\approx 1-8\pi r^2\left(\frac{\rho_0}{3}-\sum_{n=1}^N
\frac{\beta_{n}r^{n}}{n+3}  \right)\label{g11nearorigin}
\end{equation}
and
\begin{eqnarray}
\frac{\partial \nu}{\partial r}-\frac{\partial \lambda}{\partial r} &\approx & 8\pi r\left[\left(p_0-\frac{\rho_0}{3}\right)\left( 1+\frac{8\pi r^2\rho_0}{3}\right)\right.\nonumber\\ &+& \left.\sum_{n=1}^N\beta_{n} r^{n} \left( \frac{n+1}{n+3} -\frac{\partial p}{\partial\rho}\bigg |_{\rho_0} \right)\right]\label{derlog},
\end{eqnarray}
where $p_0\equiv p(\rho_0)$ is the total radial pressure at $r=0$ and the second term of Eq.' (\ref{g11nearorigin}) was assumed to be much smaller than its first one. Assume now that the dielectric properties of the matter content in the central parts of a compact system can be effectively described by those of a transparent nonlinear liquid. This is reasonable and natural since it should in principle behave similarly to a plasma, and as so it should be transparent and present low-losses for waves whose frequencies are higher than the plasma one. Thus, such a dispersive nonlinear medium would always behave as a transparent nonlinear liquid for any convenient frequency range above the plasma one such that its permittivity could be taken as only dependent upon the electric field there (practically constant in terms of the frequency). Actually, when searching for flat optical metrics, one is exactly seeking to cancel it out. From Eqs. (\ref{Er}), (\ref{g11nearorigin}) and (\ref{derlog}), one has that the norm of the electric field near the origin for a Minkowskian optical metric would behave as
\begin{equation}
E(r)\propto \left(\frac{r}{r_0} \right)^{\frac{3p_0}{\rho_0}-1}\label{Ernearorigin},
\end{equation}
being thus null there only when $p_0>\rho_0/3$. Note that any term in Eq. (\ref{derlog}) whose $r$'s power is larger than the unity is irrelevant for the issue of the regularity of the electric field.
Taking into account Eq. (\ref{mu_epsilon}) to the case under interest, one sees that its leading order term is $g_{00}(0)$, which fixes the frequency range where our analyses would be valid. Corrections to this value can be easily obtained from the integration of Eq. (\ref{derlog}).

Let us now qualitatively assess possible candidates for our formalism. As we have stated previously, any reasonable and self-consistent candidate should be such that $p_0>\rho_0/3$. For this case, due to the regularity of the electric field at $r=0$, one could safely approximate the total energy-momentum tensor there by its matter counterpart, which we shall assume to be a perfect one. In this regard, it can be checked that the well-known $\sigma-\omega-\rho$ model for dense hadronic matter within the relativistic mean-field approximation (see for instance \cite{1996csnp.book.....G, 2012NuPhA.883....1B, 2014PhRvC..89c5804R} and references therein) can satisfy such a constraint, since in its high baryon density limit (when compared to the nuclear density, $\rho_{{\rm nuc}}\approx 3\times 10^{14}\,{\rm g/cm}^3\approx 1.3\times10^{9}({\rm MeV})^4$), $p\rightarrow \rho$ from below. For instance, for the NL3 parametrization to this model \cite{1997PhRvC..55..540L}, one can verify that for $\rho\gtrsim 3.3 \rho_{{\rm nuc}}$, $p>\rho/3$ takes place. Let us see if this could also happen for systems that exhibit an innermost strange quark phase. Taking, as an example, a  MIT bag-like model to describe such a phase [$p=a(\rho -4B)$, with $a= 0.45$ and $B=(146\,{\rm MeV})^4$ (see \cite{2003LRR.....6....3S} and references therein)], one indeed has that $\rho_0\gtrsim 5.5\,\rho_{{\rm nuc}}$ also leads Eq. (\ref{Ernearorigin}) to be null at $r=0$. As we have already said, another obvious criterion that reasonable candidates for flat optical metrics must have is a nontrivial electric structure. In this regard, we stress that there is not a conclusive observational answer to that yet. It is known that the phenomenology of compact stars would evidence electric aspects to these systems only when their charge are of the order of $10^{20}$C (related to electric fields of the order of $10^{21}$V/m) \cite{2003PhRvD..68h4004R}. Naturally, our optical analyses would become meaningful for much lower charges, expected to be around $10^{8}$C or even smaller (related to fields $\lesssim 10^{10}$V/m). Nevertheless, it is very likely that phase transitions lead to the appearance of large electric fields, at least in some regions of stratified stars \cite{1996csnp.book.....G,2012NuPhA.883....1B, 2010JPhG...37g5201M}. Besides, it is also believed that strange stars should have huge electric fields near their surfaces, around $10^{20}-10^{21}$V/m \cite{2009PhRvD..80h3006N}, which also point to interior electric structures (surely of smaller magnitudes). Actually, the presence of these huge electric fields could even explain the observed high magnetic fields in compact systems \cite{2010PhRvD..82j3010P, 2011AIPC.1354...13W}, as well as masses of the order of two solar masses \cite{2009PhRvD..80h3006N}. The opposite is also naturally true: since chiefly Deutsch \cite{1955AnAp...18....1D} and Goldreich and Julian \cite{1969ApJ...157..869G}, it is well-known that high conducting magnetic stars must also present electric field structures [internal and external]. Therefore, our simple analyses suggest that stratified neutron stars with central densities of a few nuclear densities, as well as strange quark stars, could be possible candidates for Minkowskian optical geometries, at least for some of their regions, more likely the outermost ones. It also clearly shows that it is far from obvious that dense systems should be optically thick or opaque a priori. Detailed studies in this direction will be conducted elsewhere.

%%%%%%%%%%%%%%%%%%%%%%%%%%%%%%%%%%%%%%%%%%%%%%%%%%%%%%%%%%%%%%%%%%%%%%%%%%%%%
%%%%%%%%%%%%%%%%%%%%%%%%%%%%%%%%%%%%%%%%%%%%%%%%%%%%%%%%%%%%%%%%%%%%%%%%%%%%%
\section{Discussion}
\label{conc}
%%%%%%%%%%%%%%%%%%%%%%%%%%%%%%%%%%%%%%%%%%%%%%%%%%%%%%%%%%%%%%%%%%%%%%%%%%%%%
%%%%%%%%%%%%%%%%%%%%%%%%%%%%%%%%%%%%%%%%%%%%%%%%%%%%%%%%%%%%%%%%%%%%%%%%%%%%%
We have seen that the flexibility of effective metrics could have far-reaching implications. If it may be rendered or naturally becomes trivial (Minkowskian), then one (or nature) could totally cancel out light interaction with matter and fields, and light rays would propagate as if they were in vacuum. Besides, there the associated light rays would not experience anymore the phenomenon of refraction nor reflection when coming from or going to vacuum. We stress that if the medium is birefringent without the control of optical metrics, it will keep being so even when they are controlled. The reason is simply because birefringent media have two optical metrics: an ordinary and an extraordinary, and generally they could only be controlled one at a time. Only for completeness, we recall that material particles, such as electrons, would feel normally the material structure and are bound to the geometry of the background spacetime.

Light propagation analyses in material media could also be imagined for assessing their equations of state if accurate measurements and dense systems are at play. (For a proposal of assessment of the properties of spacetimes with scattering measurements, see Ref. \cite{PhysRevD.90.024022}.) One such example could be a quark-gluon plasma, which could shed a light into the issue of the microphysics of the innermost region of neutron stars, motivated by the already existent analyses for its dielectric coefficients (see \cite{2011PhRvD..84l5027L} and references therein). This would be done as follows: given the dielectric coefficients, electromagnetic fields and the optical metric of a medium, one could solve the equations for the background metric and therefore infer its equation of state. The simplest case to be analysed would be when light trajectories follow straight lines, as it is supposed to be evidenced experimentally by the lack of reflection, refraction, etc, for light when it crosses the system. For the case spherically symmetric spacetimes and electromagnetic fields are involved in Kerr media, for instance, the aforesaid analysis is summarized by Eqs.\ (\ref{h11cond2}) and (\ref{mu_epsilon}). We stress that for obtaining an unique optical metric out of the propagation of rays, in general, one should make measurements of wave aspects of light, because this is a way to differentiate between a given effective geometry and its conformal one [see $k_\mu$ in Eq. (\ref{Kintermsofraypath})]. The latter geometries are indistinguishable for the propagation of light rays alone since they follow null geodesics there (see Appendix D in Ref. \cite{1984ucp..book.....W} for further details).

Concerning Minkowskian optical metrics for compact astrophysical objects, we point out that  our analyses suggest that it would be of physical interest to investigate the situations where $p_0>\rho_0/3$. Naively speaking, it seems that stars with central densities larger than a few nuclear densities could be interesting in the above scenario. Detailed analyses would require specific systems and we leave them to be done in future works.

We generally point out that the bending of light is different for nonlinear charged black holes since light rays follow null geodesics with respect to the effective geometry, not with the background one. For this case, light rays would be naturally more appropriate to scrutinize effective theories of the electromagnetism. Properties of the background spacetime could also be assessed, but in a more indirect way, given that the effective metric felt by photons also contains the geometry felt by massive particles.

Summing up, in this work we have attempted to clarify the relevance and applicability of optical metrics in the limit of geometric optics for ray propagation (either ordinary or extraordinary). Examples concerning light propagation in nonlinear liquid media, nonlinear charged black holes and compact astrophysical systems were given for the first time to stress and evidence the possibility (which shows its flexibility, but not in a trivial way) of having simple optical metrics (such as Minkowskian) and their possible far-reaching implications in several scenarios. We plan to enter deeply into the details on these issues elsewhere.

\section{Acknowledgments}
We are indebted to the participants of the \textit{Seminario Informale} at Sapienza University for their comments apropos of this work. E.B. acknowledges the financial support provided by the CAPES-ICRANet program through the grant BEX 13956/13. J.P.P. is likewise grateful to CNPq- Conselho Nacional de Desenvolvimento Cient\'ifico e Tecnol\'ogico of the Brazilian government within the postdoctoral program ``Science without Borders''.
%%%%%%%%%%%%%%%%%%%%%%%%%%%%%%%%%%%%%%%%%%%%%%%%%%%%%%%%%%%%%%%%%%%%%%%%%


\section*{References}
\begin{thebibliography}{10}

\bibitem{1960ecm..book.....L}
L.~D. {Landau} and E.~M. {Lifshitz}.
\newblock {\em {Electrodynamics of continuous media}}.
\newblock Pergamon Press, Oxford, 1960.

\bibitem{1999poet.book.....B}
M.~{Born} and E.~{Wolf}.
\newblock {\em Principles of optics : electromagnetic theory of propagation,
  interference and diffraction of light}.
\newblock Cambridge University Press, Cambridge, 1999.

\bibitem{1970JMP....11..941B}
G.~{Boillat}.
\newblock {Nonlinear Electrodynamics: Lagrangians and Equations of Motion}.
\newblock {\em Journal of Mathematical Physics}, 11:941--951, March 1970.

\bibitem{2000PhRvD..61d5001N}
M.~{Novello}, V.~A. {de Lorenci}, J.~M. {Salim}, and R.~{Klippert}.
\newblock {Geometrical aspects of light propagation in nonlinear
  electrodynamics}.
\newblock {\em \prd}, 61(4):045001, February 2000.

\bibitem{2000PhRvA..62a2111L}
U.~{Leonhardt}.
\newblock {Space-time geometry of quantum dielectrics}.
\newblock {\em \pra}, 62(1):012111, July 2000.

\bibitem{2001PhRvD..63j3516N}
M.~{Novello}, J.~M. {Salim}, V.~A. {de Lorenci}, and E.~{Elbaz}.
\newblock {Nonlinear electrodynamics can generate a closed spacelike path for
  photons}.
\newblock {\em \prd}, 63(10):103516, May 2001.

\bibitem{2002PhRvD..65f3501D}
V.~A. {de Lorenci}, R.~{Klippert}, M.~{Novello}, and J.~M. {Salim}.
\newblock {Nonlinear electrodynamics and FRW cosmology}.
\newblock {\em \prd}, 65(6):063501, March 2002.

\bibitem{2002PhRvD..66b4042O}
Y.~N. {Obukhov} and G.~F. {Rubilar}.
\newblock {Fresnel analysis of wave propagation in nonlinear electrodynamics}.
\newblock {\em \prd}, 66(2):024042, July 2002.

\bibitem{2007CQGra..24.3021N}
M.~{Novello}, E.~{Goulart}, J.~M. {Salim}, and S.~E. {Perez Bergliaffa}.
\newblock {Cosmological effects of nonlinear electrodynamics}.
\newblock {\em Classical and Quantum Gravity}, 24:3021--3036, June 2007.

\bibitem{1923AnP...377..421G}
W.~{Gordon}.
\newblock {Zur Lichtfortpflanzung nach der Relativit{\"a}tstheorie}.
\newblock {\em Annalen der Physik}, 377:421--456, 1923.

\bibitem{1999PhLB..458..466O}
Y.~N. {Obukhov} and F.~W. {Hehl}.
\newblock {Spacetime metric from linear electrodynamics}.
\newblock {\em Physics Letters B}, 458:466--470, July 1999.

\bibitem{2000PhRvD..62d4050O}
Y.~N. {Obukhov}, T.~{Fukui}, and G.~F. {Rubilar}.
\newblock {Wave propagation in linear electrodynamics}.
\newblock {\em \prd}, 62(4):044050, August 2000.

\bibitem{2003CQGra..20..859N}
M.~{Novello}, S.~{Perez Bergliaffa}, J.~{Salim}, V.~A. {DeLorenci}, and
  R.~{Klippert}.
\newblock {Analogue black holes in flowing dielectrics}.
\newblock {\em Classical and Quantum Gravity}, 20:859--871, March 2003.

\bibitem{2001PhLB..512..417D}
V.~A. {De Lorenci} and M.~A. {Souza}.
\newblock {Electromagnetic wave propagation inside a material medium: an
  effective geometry interpretation}.
\newblock {\em Physics Letters B}, 512:417--422, July 2001.

\bibitem{2002PhRvE..65b6612D}
V.~A. {de Lorenci}.
\newblock {Effective geometry for light traveling in material media}.
\newblock {\em \pre}, 65(2):026612, February 2002.

\bibitem{2002PhRvD..65f4027D}
V.~A. {de Lorenci} and R.~{Klippert}.
\newblock {Analogue gravity from electrodynamics in nonlinear media}.
\newblock {\em \prd}, 65(6):064027, March 2002.

\bibitem{2003PhRvD..68f1502D}
V.~A. {de Lorenci}, R.~{Klippert}, and Y.~N. {Obukhov}.
\newblock {Optical black holes in moving dielectrics}.
\newblock {\em \prd}, 68(6):061502, September 2003.

\bibitem{2010NJPh...12i5021C}
S.~L. {Cacciatori}, F.~{Belgiorno}, V.~{Gorini}, G.~{Ortenzi}, L.~{Rizzi},
  V.~G. {Sala}, and D.~{Faccio}.
\newblock {Spacetime geometries and light trapping in travelling refractive
  index perturbations}.
\newblock {\em New Journal of Physics}, 12(9):095021, September 2010.

\bibitem{2011LRR....14....3B}
C.~{Barcel{\'o}}, S.~{Liberati}, and M.~{Visser}.
\newblock {Analogue Gravity}.
\newblock {\em Living Reviews in Relativity}, 14:3, May 2011.

\bibitem{1981PhRvL..46.1351U}
W.~G. {Unruh}.
\newblock {Experimental black-hole evaporation}.
\newblock {\em Physical Review Letters}, 46:1351--1353, May 1981.

\bibitem{2011CQGra..28n5022N}
M.~{Novello} and E.~{Goulart}.
\newblock {Beyond analog gravity: the case of exceptional dynamics}.
\newblock {\em Classical and Quantum Gravity}, 28(14):145022, July 2011.

\bibitem{2008arXiv0805.4778L}
U.~{Leonhardt} and T.~G. {Philbin}.
\newblock {Transformation Optics and the Geometry of Light}.
\newblock {\em ArXiv e-prints}, May 2008.

\bibitem{2003IJMPD..12..649V}
M.~{Visser}.
\newblock {Essential and Inessential Features of Hawking Radiation}.
\newblock {\em International Journal of Modern Physics D}, 12:649--661, 2003.

\bibitem{2010PhRvL.105t3901B}
F.~{Belgiorno}, S.~L. {Cacciatori}, M.~{Clerici}, V.~{Gorini}, G.~{Ortenzi},
  L.~{Rizzi}, E.~{Rubino}, V.~G. {Sala}, and D.~{Faccio}.
\newblock {Hawking Radiation from Ultrashort Laser Pulse Filaments}.
\newblock {\em Physical Review Letters}, 105(20):203901, November 2010.

\bibitem{2011PhRvL.107n9401S}
R.~{Sch{\"u}tzhold} and W.~G. {Unruh}.
\newblock {Comment on ``Hawking Radiation from Ultrashort Laser Pulse
  Filaments''}.
\newblock {\em Physical Review Letters}, 107(14):149401, September 2011.

\bibitem{2008Sci...319.1367P}
Thomas~G. {Philbin}, Chris {Kuklewicz}, Scott {Robertson}, Stepenn {Hill},
  Friedrich {K{\"o}nig}, and U.~{Leonhardt}.
\newblock {Fiber-Optical Analog of the Event Horizon}.
\newblock {\em Science}, 319:1367, March 2008.

\bibitem{2008NJPh...10e3015R}
G.~{Rousseaux}, C.~{Mathis}, P.~{Ma{\"i}ssa}, T.~G. {Philbin}, and
  U.~{Leonhardt}.
\newblock {Observation of negative-frequency waves in a water tank: a classical
  analogue to the Hawking effect?}
\newblock {\em New Journal of Physics}, 10(5):053015, May 2008.

\bibitem{PhysRevLett.106.021302}
Silke Weinfurtner, Edmund~W. Tedford, Matthew C.~J. Penrice, William~G. Unruh,
  and Gregory~A. Lawrence.
\newblock Measurement of stimulated hawking emission in an analogue system.
\newblock {\em Phys. Rev. Lett.}, 106:021302, Jan 2011.

\bibitem{2012CQGra..29v4009F}
D.~{Faccio}, T.~{Arane}, M.~{Lamperti}, and U.~{Leonhardt}.
\newblock {Optical black hole lasers}.
\newblock {\em Classical and Quantum Gravity}, 29(22):224009, November 2012.

\bibitem{2010PhRvA..82e3602K}
Y.~{Kurita}, M.~{Kobayashi}, H.~{Ishihara}, and M.~{Tsubota}.
\newblock {Particle creation in Bose-Einstein condensates: Theoretical
  formulation based on conserving gapless mean-field theory}.
\newblock {\em \pra}, 82(5):053602, November 2010.

\bibitem{2014PhRvD..90j4015B}
A.~{Belenchia}, S.~{Liberati}, and A.~{Mohd}.
\newblock {Emergent gravitational dynamics in a relativistic Bose-Einstein
  condensate}.
\newblock {\em \prd}, 90(10):104015, November 2014.

\bibitem{2007arXiv0712.0427V}
M.~{Visser} and S.~{Weinfurtner}.
\newblock {Analogue spacetimes: Toy models for ''quantum gravity''}.
\newblock {\em ArXiv e-prints}, December 2007.

\bibitem{1975ctf..book.....L}
L.~D. {Landau} and E.~M. {Lifshitz}.
\newblock {\em {The classical theory of fields}}.
\newblock Pergamon Press, Oxford, 1975.

\bibitem{2000PhRvL..84.4184S}
D.~R. {Smith}, W.~J. {Padilla}, D.~C. {Vier}, S.~C. {Nemat-Nasser}, and
  S.~{Schultz}.
\newblock {Composite Medium with Simultaneously Negative Permeability and
  Permittivity}.
\newblock {\em Physical Review Letters}, 84:4184--4187, May 2000.

\bibitem{2001Sci...292...77S}
R.~A. {Shelby}, D.~R. {Smith}, and S.~{Schultz}.
\newblock {Experimental Verification of a Negative Index of Refraction}.
\newblock {\em Science}, 292:77--79, April 2001.


\bibitem{2010PhRvL.105f7402S}
I.~I. {Smolyaninov} and E.~E. {Narimanov}.
\newblock {Metric Signature Transitions in Optical Metamaterials}.
\newblock {\em Physical Review Letters}, 105(6):067402, August 2010.

\bibitem{2008PhRvD..78d5015D}
V.~A. {de Lorenci} and G.~P. {Goulart}.
\newblock {Magnetoelectric birefringence revisited}.
\newblock {\em \prd}, 78(4):045015, August 2008.

\bibitem{2013LRR....16....5A}
G.~{Amelino-Camelia}.
\newblock {Quantum-Spacetime Phenomenology}.
\newblock {\em Living Reviews in Relativity}, 16, June 2013.

\bibitem{1993PhRvD..48.3641B}
J.~D. {Bekenstein}.
\newblock {Relation between physical and gravitational geometry}.
\newblock {\em \prd}, 48:3641--3647, October 1993.

\bibitem{2010opme.book.....C}
W.~{Cai} and V.~{Shalaev}.
\newblock {\em Optical Metamaterials: Fundamentals and Applications}.
\newblock Springer-Verlag, New York, 2010.


\bibitem{1934RSPSA.144..425B}
M.~{Born} and L.~{Infeld}.
\newblock {Foundations of the New Field Theory}.
\newblock {\em Royal Society of London Proceedings Series A}, 144:425--451,
  March 1934.

\bibitem{2014PhRvA..89d3822D}
V.~A. {De Lorenci} and J.~P. {Pereira}.
\newblock {One-way propagation of light in Born-Infeld-like metamaterials}.
\newblock {\em \pra}, 89(4):043822, April 2014.

\bibitem{2014PhLB..734..396P}
J.~P. {Pereira}, H.~J. {Mosquera Cuesta}, J.~A. {Rueda}, and R.~{Ruffini}.
\newblock {On the black hole mass decomposition in nonlinear electrodynamics}.
\newblock {\em Physics Letters B}, 734:396--402, June 2014.

\bibitem{2001PhRvD..63h3511N}
M.~{Novello} and J.~M. {Salim}.
\newblock {Effective electromagnetic geometry}.
\newblock {\em \prd}, 63(8):083511, April 2001.

\bibitem{2012PhRvD..86l4024N}
M.~{Novello} and E.~{Bittencourt}.
\newblock {Gordon metric revisited}.
\newblock {\em \prd}, 86(12):124024, December 2012.

\bibitem{1999PhRvA..60.4301L}
U.~{Leonhardt} and P.~{Piwnicki}.
\newblock {Optics of nonuniformly moving media}.
\newblock {\em \pra}, 60:4301--4312, December 1999.

\bibitem{1996csnp.book.....G}
N.~{Glendenning}.
\newblock {\em Compact Stars.~ Nuclear Physics, Particle Physics and General
  Relativity}.
\newblock Springer-Verlag, New York, 1996.

\bibitem{2012NuPhA.883....1B}
R.~{Belvedere}, D.~{Pugliese}, J.~A. {Rueda}, R.~{Ruffini}, and S.-S. {Xue}.
\newblock {Neutron star equilibrium configurations within a fully relativistic
  theory with strong, weak, electromagnetic, and gravitational interactions}.
\newblock {\em Nuclear Physics A}, 883:1--24, June 2012.

\bibitem{2014PhRvC..89c5804R}
J.~A. {Rueda}, R.~{Ruffini}, Y.-B. {Wu}, and S.-S. {Xue}.
\newblock {Surface tension of the core-crust interface of neutron stars with
  global charge neutrality}.
\newblock {\em \prc}, 89(3):035804, March 2014.

\bibitem{1997PhRvC..55..540L}
G.~A. {Lalazissis}, J.~{K{\"o}nig}, and P.~{Ring}.
\newblock {New parametrization for the Lagrangian density of relativistic mean
  field theory}.
\newblock {\em \prc}, 55:540--543, January 1997.

\bibitem{2003LRR.....6....3S}
N.~{Stergioulas}.
\newblock {Rotating Stars in Relativity}.
\newblock {\em Living Reviews in Relativity}, 6, June 2003.

\bibitem{2003PhRvD..68h4004R}
S.~{Ray}, A.~L. {Esp{\'{\i}}ndola}, M.~{Malheiro}, J.~P. {Lemos}, and V.~T.
  {Zanchin}.
\newblock {Electrically charged compact stars and formation of charged black
  holes}.
\newblock {\em \prd}, 68(8):084004, October 2003.

\bibitem{2010JPhG...37g5201M}
I.~N. {Mishustin}, C.~{Ebel}, and W.~{Greiner}.
\newblock {Strong electric fields induced on a sharp stellar boundary}.
\newblock {\em Journal of Physics G Nuclear Physics}, 37(7):075201, July 2010.

\bibitem{2009PhRvD..80h3006N}
R.~P. {Negreiros}, F.~{Weber}, M.~{Malheiro}, and V.~{Usov}.
\newblock {Electrically charged strange quark stars}.
\newblock {\em \prd}, 80(8):083006, October 2009.

\bibitem{2010PhRvD..82j3010P}
R.~{Pican{\c c}o Negreiros}, I.~N. {Mishustin}, S.~{Schramm}, and F.~{Weber}.
\newblock {Properties of bare strange stars associated with surface electric
  fields}.
\newblock {\em \prd}, 82(10):103010, November 2010.

\bibitem{2011AIPC.1354...13W}
F.~{Weber} and R.~{Negreiros}.
\newblock {QCD in Neutron Stars and Strange Stars}.
\newblock In A.~{K{\i}z{\i}lers{\"u}} and A.~W. {Thomas}, editors, {\em
  American Institute of Physics Conference Series}, volume 1354 of {\em
  American Institute of Physics Conference Series}, pages 13--18, May 2011.

\bibitem{1955AnAp...18....1D}
A.~J. {Deutsch}.
\newblock {The electromagnetic field of an idealized star in rigid rotation in
  vacuo}.
\newblock {\em Annales d'Astrophysique}, 18:1, January 1955.

\bibitem{1969ApJ...157..869G}
P.~{Goldreich} and W.~H. {Julian}.
\newblock {Pulsar Electrodynamics}.
\newblock {\em \apj}, 157:869, August 1969.


\bibitem{PhysRevD.90.024022}
Jason Doukas, Luke Westwood, Daniele Faccio, Andrea Di~Falco, and Ivette
  Fuentes.
\newblock Gravitational parameter estimation in a waveguide.
\newblock {\em Phys. Rev. D}, 90:024022, Jul 2014.

\bibitem{2011PhRvD..84l5027L}
J.~{Liu}, M.~J. {Luo}, Q.~{Wang}, and H.-J. {Xu}.
\newblock {Refractive index of light in the quark-gluon plasma with the
  hard-thermal-loop perturbation theory}.
\newblock {\em \prd}, 84(12):125027, December 2011.

\bibitem{1984ucp..book.....W}
R.~M. {Wald}.
\newblock {\em General relativity}.
\newblock University of Chicago Press, Chicago, 1984.

\end{thebibliography}
\end{document}